\documentstyle[aps]{revtex}
\input{epsf}

\newcommand{\bq}{\begin{equation}}
\newcommand{\ee}{\end{equation}}
\newcommand{\bea}{\begin{eqnarray}}
\newcommand{\eea}{\end{eqnarray}}
\newcommand{\fr}[2]{\frac{#1}{#2}}

\newcommand{\mbf}{\mathbf}

\begin{document}
\draft
\title{Instability of a trapped ultracold Fermi gas 
with attractive interactions: quantum effects}
\author{X. Leyronas and R. Combescot}
\address{ 
Laboratoire de Physique Statistique, Ecole Normale Sup\'erieure*, 
24 rue Lhomond, 75231 Paris Cedex 05, France\\ 
}
\date{\today}
\maketitle
\begin{abstract}
We consider the possible mechanical instability of an ultracold
Fermi gas due to the attractive interactions between fermions
of different species. We investigate how the instability, predicted
by a mean field calculation, is modified
when the gas is trapped in a harmonic potential and 
quantum effects are included.
\end{abstract}
%\narrowtext

\section{Introduction}

The field of ultracold Fermi gases has recently seen a very strong
development.
After reaching the degenerate regime, experiments have dealt with mixtures
of two hyperfine levels \cite{dieck,ohara,cubi} . This has been done mostly in the vicinity of a 
Feshbach resonance, which allows quite conveniently to adjust the effective
interaction. Indeed by changing the magnetic field, one can start with a small
negative scattering length, make it more negative, let it have a jump
from $- \infty $ to $+ \infty $ at the resonance, and then have it decrease
to small positive values. A major purpose of these experiments is to look
for the BCS transition, which should occur in particular
for negative scattering length
$a$. Naturally one expects the critical temperature to be higher for larger
$ |a|$, since this corresponds to a stronger attractive interaction. 

On the other hand one may expect that this overall attractive interaction
may give rise also to a collapse instability. This would be 
similar to the one very much studied in Bose-Einstein condensates \cite{ueda},
where this instability prohibits the formation of condensates with a large
number of atoms. It is reasonable to think that similarly the BCS instability
would be in competition with a collapse.
The first work where the BCS transition for ultracold
atoms was explicitely considered \cite{stoofal} studied also this collapse
within the mean field approximation, considering in particular the
dependence on the ratio between the number of atoms in the two atomic
population expected to form Cooper pairs. The vicinity of this collapse
would also be particularly favorable \cite{rc} to the BCS transition.

More specifically let us consider only the case of two equal 
populations of atoms corresponding to two different hyperfine levels,
and restrict ourselves to the $ T = 0 $ case. 
A scattering length $a$ corresponds to an effective
interaction constant $ g = 4 \pi \hbar ^{2}a / m $ between unlike
atoms, giving rise
to a mean field contribution $ g n/2 $ to the chemical potential
$ \mu (n)$ where $n$ is the total atomic density. Including kinetic
energy the total chemical potential is given in this approximation
by $ \mu (n) = \hbar ^{2} k _{F}^{2}/2m - | g | n/2 $
with $ 3 \pi ^{2} n = k _{F}^{3}$. The instability is obtained from
$\partial \mu  / \partial n  = 0$ which gives for the critical
density $ \lambda
\equiv  2 k _{F}| a | / \pi  = 1 $.

Quite remarkably recent experiments \cite{dieck,ohara,cubi,gehm}
on mixtures of fermionic atoms 
with two different hyperfine states have not observed this instability
when they have been through the Feshbach resonance, where the scattering
length becomes infinitely large. Since these experiments have reached
quite low temperatures it is very unlikely that they missed this
transition because they did not go at low enough $T$.
On one hand the absence of this
instability is quite fortunate since it allows experiments to reach all
the possible range of scattering lengths, without any limitation.
In particular this gives the possibility to produce molecules, which
has been extremely fruitful very recently.
Nevertheless the failure of the mean field calculation \cite{stoofal}
is somewhat striking since, although there is no reason to believe
mean field to be quantitatively correct, it gives quite often
reasonable qualitative estimates. Note however that there is nothing
basically wrong in the idea that a collapse instability should exist
since it has been indeed observed experimentally
\cite{modu} in a fermion-boson mixture
with an effective attraction between fermions and bosons. 

In this paper we investigate a possible source for this disagreement
between this simple mean field estimate \cite{rothfeld} and experiments. 
An indication in this direction can be found in
the results of an investigation of the hydrodynamic modes in harmonic
trap \cite{rcxl} . The hydrodynamic framework implies 
that the quantum length scale
does not appear in the study, except indirectly in the equation of
state of the dense gas. The surprising result is that no mode with
zero frequency is actually found, even when the system reaches the
density for collapse at the center of the trap. All the modes are
found to have non zero frequencies, whereas one would expect the
collapse instability to manifest itself by the appearance of a mode
with zero frequency, as it is found for a Bose condensate. 
Indeed one can show, from the starting equations, that it is a general feature of these hydrodynamic modes 
to have non-zero frequencies \cite{moden0}.
This result may suggest that, in a trap, the
collapse may be missed in some way. 

In order to explore this problem and go beyond the macroscopic scale
by including quantum effects,
we perform in this paper a semiclassical
microscopic calculation to find if there is a zero frequency mode.
Since experiments have already reached very low temperatures, and that
this is the most favorable situation for appearance of the
instability, we restrict our exploration to the $T=0$ case.
We find indeed that there is one.
The lack of zero frequency mode in the hydrodynamic framework
has actually a simple physical
explanation. The gas becomes unstable when the density at the center
of the trap reaches the critical density. However the gas in all
the other parts of the trap has a density lower than this critical
density. So the overall system is, so to speak, not soft enough to
a have a zero frequency mode. When one goes to a microscopic calculation,
the unstable region, which had at the macroscopic scale a zero extension,
gets a finite extension of order of the microscopic scale. In this way
this region can produce a strong enough softening and give rise to
a zero frequency mode, which we obtain below explicitely.

In the next section we begin our investigation by making a simple 
RPA approximation to find
the microscopic effect of interactions. Our treatment takes 
into account the modification
of the density distribution due to the interactions.
This allows us to find out explicitely the important features in this
problem, and in particular to point out the relevant length scales.
We can then generalize
our approach by getting rid of the RPA approximation, and show how
to deal with the problem on the quite general grounds of Fermi liquid
theory. We find indeed that the threshold for instability is
modified by the trap and obtain explicitely this modification, as well
as the shape of the mode responsible for the instability. However,
although this modification could be sizeable for the very anisotropic
traps used in some experiments, it appears unlikely that it is
responsible for the overall disappearance of the instability, as it
is observed experimentally.

\section{Theoretical treatment of the instability in a trap}

We consider now the above atomic gas of fermions with mass
$m$, in an isotropic harmonic trap of frequency $\Omega$, giving rise
to the harmonic potential $ V( {\bf x } ) = \frac{1}{2} m \Omega ^{2} r^{2}$
with $r^{2}={\bf x }^2$. 
An (undamped) eigenmode corresponds to an infinite response of the
system excited at the frequency of the mode. Since we are interested
in a zero frequency mode \cite{instabmod},
we have to consider in addition 
a static perturbation $\delta V({\mbf x})$, 
which will induce a static density fluctuation 
$\delta \rho({\mbf x})$. The collapse instability 
will correspond to a divergent density fluctuation. 
The linear response theory gives\cite{fetwale}:
\bea
\delta\rho({\mbf x})&=&\int\,d{\mbf x'}\,\Pi({\mbf x}, {\mbf x'})\delta V({\mbf x'})
\eea
where $\Pi({\mbf x},{\mbf x'})=-\frac{i}{\hbar}\int\,dt\,\theta(t)
\langle \left[\hat{\rho}({\mbf x},t),\hat{\rho}({\mbf x'},0)\right]\rangle$
is the zero frequency density-density (retarded) response function. 
Therefore an instability corresponds to a divergent eigenvalue
of the kernel operator $\Pi({\mbf x},{\mbf x'})$, which is real symmetric. 
For interacting particles, one has to make use of some kind of
approximation to calculate $\Pi$. 
The simplest one, which will reduce as we will see to the mean field result
for an infinitely wide trap (in which case the system would be homogeneous)
is the Random Phase Approximation (RPA) and this is the one
we will use here. As it is well known, 
it is equivalent to sum up an infinite series of bubble diagrams.
Since the fermions we deal with interact with a very short range
atomic size potential, we can use a contact potential $U( {\bf x } )
= g \delta  ( {\bf x } )$ , with $ g = 4 \pi \hbar ^{2}a / m $,
for the interaction potential, where
$g < 0$ since we consider an attractive interaction.
In this case the RPA for our trapped fermions reads:
\bea
\Pi({\mbf x},{\mbf x'})&=&\Pi^{0}({\mbf x},{\mbf x'}) 
+g\int\,d{\mbf x_1}\,\Pi^{0}({\mbf x},{\mbf x_1})\,\Pi({\mbf x_1},{\mbf x'})
\label{eqrpa}
%\Pi({\mbf r},{\mbf r'})&=&\Pi^{0}({\mbf r},{\mbf r'}) 
%+\int\,d{\mbf r_1}\,d{\mbf r_2}\,\Pi^{0}({\mbf r},{\mbf r_1}) g({\mbf r_1}-{\mbf r_2})\,\Pi({\mbf r_2},{\mbf r'})
\eea
This Eq. (\ref{eqrpa}) reads formally $\Pi=\Pi^{0}+g\,\Pi^0\,\Pi$, 
and its formal solution is
$\Pi=(1-g\Pi^{0})^{-1}\Pi^0$. 
This shows that the eigenvectors of $\Pi$ and 
$\Pi^0$ are the same, and that the instability we look for appears when 
the smallest (negative) eigenvalue of $\Pi^0$ equals $1/g$. 
The general eigenvalue equation for $\Pi^0$ (eigenvalue $\alpha$, eigenvector 
$\varphi$) reads:
\bea
\int d{\mbf x'}\,\Pi^0({\mbf x},{\mbf x'})\,\varphi({\mbf x'})&=&
\alpha\,\varphi({\mbf x})\label{eqmode1}
\eea 
It is then convenient to introduce the Wigner transform 
\mbox{$\Pi^{0}_{W}({\mbf q},{\mbf R})=\int\,d{\mbf r}\,
e^{-i{\mbf q}.{\mbf r}}\Pi^0({\mbf R}+{\mbf r}/2,{\mbf R} -{\mbf r}/2)$}. 
Eq.(\ref{eqmode1}) then becomes:
\bea
\int \frac{d{\mbf r}d{\mbf q}}{(2\pi)^3}\,e^{i{\mbf q}.{\mbf r}}\,
\Pi^{0}_{W}({\mbf q},{\mbf x}-{\mbf r}/2)\,\varphi({\mbf x}-{\mbf r})&=&
\alpha\,\varphi({\mbf x})\label{eqmodeW}
\eea 

Then we use the fact that there is a large number of trapped 
particles or equivalently that the chemical potential is much larger
than the level spacing $\hbar \Omega$ . This allows to make use of a 
semiclassical treatment by considering that the trapping potential
is slowly varying. In this case we can use for 
$\Pi^{0}_{W}({\mbf q},{\mbf R})$ its homogeneous value, evaluated with
the local value of the particle density.
We therefore make the approximation 
$\Pi^{0}_{W}({\mbf q},{\mbf R})\approx \Pi^{0}({\mbf q})$, where  
$\Pi^{0}({\mbf q})$ is the response function of the homogeneous
system with a Fermi wave vector $k_{F}(R)$. The local Fermi wave vector is 
related to the equilibrium density of the cloud $n(R)=k_{F}^3/3\pi^2$,
determined by the equation:
\bea
\mu(n)+1/2\,m\,\Omega^2\, r^2=\tilde{\mu}
\eea
where $\mu(n)$ is the chemical potential and $\tilde{\mu}$ is the overall chemical
potential. 

Our next step is to take advantage of the length scale $d$
we expect physically for the instability mode we are interested in.
Clearly the instability will occur at the center of the trap since
this is where the local particle density is highest. On the other
hand this mode is a collective phenomenon involving a large number
of particles, so it must occur over a typical scale large compared
to the interparticle distance, which is itself of order $k_F^{-1}$. This leads
us to look for a mode which satisfies
$k_{F}^{-1}\ll d\ll R_{0}$. This relation implies that the typical
wavevectors entering the Fourier expansion of $\varphi({\mbf x})$
are small compared to $k_F$. From Eq.(\ref{eqmodeW}) it is then seen that
the wavevector ${\mbf q}$ in $\Pi^{0}_{W}({\mbf q},{\mbf R})$ must
also be small compared to $k_F$. This
allows us to expand $\Pi^0$ in powers of ${\mbf q}$. Since
$\Pi^{0}({\mbf q})$ is just the free particle response function,
we have \cite{fetwale} explicitely 
$\Pi^{0}({\mbf q})\approx\,
-\frac{1}{2\pi^2}m\,k_{F}(1-\frac{1}{12}q^2/k_{F}^2)$. 
When we insert this expression in Eq. (\ref{eqmodeW}) and perform the 
integrals, we find
the following second order partial differential equation for 
the density fluctuation $\varphi$ corresponding to the instability mode: 
\bea
\Delta\varphi +k_{F}\nabla(\frac{1}{k_{F}}).\nabla\varphi 
+\frac{1}{4}k_{F}\varphi\Delta(\frac{1}{k_F})
+12(k_{F}^2+2\pi^2 \frac{k_{F}}{m}\,\alpha)\varphi&=&0\label{eqdiff1}
\eea

If we consider now the order of magnitude of the three first terms of 
Eq.(\ref{eqdiff1}), we notice that the second and the third term contain derivatives of $k_{F}$, 
while the first one contains only derivatives of $\varphi$. 
Since the length scale $R_0$ for the variations of $k_F$ is much 
larger than the length scale $d$ for the variations of $\varphi$, the second and third terms are negligible compared to the 
first one. 

In order to solve explicitely Eq.(\ref{eqdiff1}) we consider as a first 
step, in the following subsection, the simple case where the modification
of the density distribution due to the interactions is not taken into
account. This will allow us to see clearly the relevant length scales.
This simple case is equivalent to assume the free particle relation
$\mu(n)=\hbar^2\,k_{F}^2/2m$ for the equation of state.
We will then take consistently into account interactions in $k_{F}(R)$
in the next subsection.

\subsection{Simple case}

In this simple case we have merely \cite{bruunclark}:
\bea
k_{F}(R)&=&k^{0}_{F}(1-R^2/R_0^2)^{1/2}\label{eqkf}
\eea
where the Thomas-Fermi cloud radius $R_0$ and the maximum  Fermi wave
vector $k^0_{F}$ are related by \cite{bruunclark}
$\hbar^2\,(k^0_{F})^2 /2m\,=\,(1/2)\,m\,\Omega^2\,R_{0}^2$. Both are
directly related to the particle number $N$ in the trap. 
Coming back to Eq. (\ref{eqdiff1}), since again $d\ll R_{0}$, $k_{F}(x)$ is a slowly varying function in
the considered domain of  $x\equiv \|{\mbf x}\|\sim d$. 
We can therefore expand $k_{F}(x)$ up to second order
around $x=0$, using Eq.(\ref{eqkf}). 
Setting the eigenvalue $\alpha=1/g$ at the instability, we can  introduce the coupling constant 
$\lambda=-\frac{1}{2\pi^2}mk_{F}^0 g = -\frac{2}{\pi}k_{F}^0 a$, and we find:
\bea
\Delta\varphi+12\left[
(k_{F}^0)^{2}(1-1/\lambda)-(k_{F}^0/R_0)^2(1-1/2\lambda)x^2\right]\varphi &=&0\label{eqdiffoh}
\eea 
Going up to fourth order in the expansion of $k_{F}(x)$ would give a term of order $(k_{F}^{0})^2/(R_{0})^4\,x^4$,
which is smaller than the $x^2$ term in Eq.(\ref{eqdiffoh}) 
by a factor $(x/R_{0})^2\ll 1$. It is therefore justified
to stop the small $x$ expansion to second order to get Eq.(\ref{eqdiffoh}) from
Eq.(\ref{eqdiff1}).

Now Eq.(\ref{eqdiffoh}) is the Schr\"odinger equation for the 3D harmonic oscillator of frequency $\omega$ 
for a state of energy $E$, 
provided we set $\hbar=m=1$, together with:
\bea 
E=6(k_{F}^0)^{2}(1-\fr{1}{\lambda}) 
\label{energlam}
\eea
and
\bea
\omega^2=12 (k_{F}^0/R_0)^2(1-\fr{1}{2\lambda})
\label{eqomega}
\eea
Since the instability mode we are looking for is naturally localized in the
center of the trap, we are looking for the bound states of this harmonic
oscillator. The critical value of the coupling constant $ \lambda  $ 
is directly obtained from the energy $E$ of the oscillator 
by Eq.(\ref{energlam}). Since the first
instability will occur for the smallest value of the coupling constant,
we are looking from this equation for the smallest possible value of the energy.
We check that we recover properly the homogeneous case by taking the 
limit $ R_0 \rightarrow \infty$. Indeed in this case the harmonic 
oscillator frequency $\omega$ goes to zero, which implies that all the bound
states energies go to zero. From Eq.(\ref{energlam}) this gives $ \lambda  = 1$
as expected. Coming back to the trapped case we will find the lower energy
among the isotropic $s-$wave states. We have for these states
the quantization relation $E=\omega(2n+3/2)$, 
with $n=0,1,\cdots$. For a given $n$, this determines $\lambda$ at the instability. 
The smallest value of $\lambda$ is obtained for $n=0$, corresponding to the gaussian mode 
$\varphi({\mbf x})=e^{-\frac{1}{2}\omega\,x^2}$. This yields
$1-1/\lambda=\sqrt{3/4(1-1/2\lambda)}(R_{0}k_{F}^0)^{-1}\ll 1$. This
result shows that we have $ \lambda  \simeq 1$. Therefore to first
order in $ 1/R_{0}k_{F}^0$ we have at the instability:
\bea
\lambda - 1 &\approx& \sqrt{\fr{3}{8}} \, \fr{1}{R_{0}k_{F}^0} \label{eqlambda}
\eea

Naturally we have to check the consistency of our calculation by looking
at the size $d$ of the instability mode. Since we have 
$\lambda \simeq 1$, we can just set $\lambda= 1$
in the equation Eq.(\ref{eqomega}) for $\omega^2$. 
We get $\omega=\sqrt{6}k_{F}^{0}/R_0$
leading to a gaussian mode 
$\varphi({\mbf x})=\exp(-\frac{1}{2}x^2/d^2)$
of width 
$d \equiv \omega^{-1/2}=6^{-1/4}(R_{0}/k_{F}^{0})^{1/2}=6^{-1/4}\sqrt{\hbar/m\Omega}$.
Except for the numerical coefficient this is the size of the single 
particle ground state in the harmonic trapping
potential. This result for the width of the mode is completely consistent 
with our starting hypotheses $k_{F}^{-1}\ll d\ll R_{0}$. 

\subsection{Self-consistent calculation}

Now, since we have the quantitative situation under control, we come
back to a consistent description of the density distribution
in the atomic cloud, taking interactions into account
in the calculation of the Fermi wave vector. Taking for the chemical potential 
the Hartree approximation, we have $\mu(n)=\hbar^2 k_{F}^2/2m - | g | n/2$ with 
$n=k_{F}^3 /3\pi^2$. 
  When $\lambda=1$, the static compressibility $(n\partial\mu/\partial n)^{-1}$ 
diverges at the center of the trap. As a result when we take the derivative
of the chemical potential with respect to the Fermi wavevector, we find
that it is zero. This implies that $k_{F}(R)$ is a linear function of $R$
close to the trap center, instead of being quadratic as in Eq.(\ref{eqkf}).
As we have seen in the preceding subsection, we need to know $k_{F}(R)$
only close to the center. This is easily done and one finds:
\bea
\fr{k_{F}(R)}{k^{0}_{F}} &\approx& 1 -\frac{1}{\sqrt{3}} \fr{R}{R_{0}},\label{eqkf2} 
\eea
where $R_{0}$ is again the Thomas-Fermi radius of the atomic cloud.
But it is now related to the Fermi wavevector at the center $k^{0}_{F}$
by $\hbar k^{0}_{F}= \sqrt{3} m \Omega  R_{0}$, instead of 
$\hbar k^{0}_{F}=  m \Omega  R_{0}$ as above.
Following the same procedure as before, we insert 
this expression for the density into Eq.(\ref{eqdiff1}).
Keeping only the dominant terms, we get:
\bea
\Delta \varphi +12 (k^{0}_{F})^2\,(1- \frac{1}{\lambda  }
 -\frac{1}{\sqrt{3}} \frac{x}{R_{0}})\varphi
&=& 0 \label{eqdiffairy}
\eea
For s-wave solutions this equation can be reduced to Airy function 
differential equation:
\bea
\psi''(y)\,=\,(y-y_{0})\,\psi(y)
\eea
provided that we rescale
the position $x$ according to $x=D\,y$, with 
the new length scale $D$ given by
$D/R_{0}=(4\sqrt{3}(k^{0}_{F}\,R_{0})^2))^{-1/3}$, and
we introduce the new function $\psi=x\,\varphi$. 
We find $y_{0}=12 (k^{0}_{F}\,D)^2 \,(1-1/\lambda)$.
We note that the power law dependence of $D/R_0$ on $k^{0}_{F}R_{0}$
is slightly different from the one we have found for $d/R_0$.
Nevertheless our starting hypotheses $k_{F}^{-1}\ll d\ll R_{0}$
are still satisfied.
The boundary conditions $\psi(0)=0$ and $\psi(+\infty)=0$ impose that
$\psi(y)=Ai(y-y_{0})$ where $y_0\approx 2.3$ is the first zero of the Airy function.
We finally get for the coupling constant at which the instability arises:
\bea
\lambda  - 1  &\approx& \frac{y_{0}}{(6 k^{0}_{F}\,R_{0})^{2/3}}  
\label{eqcoll2}
\eea
This result is similar to the one we have found above Eq.(\ref{eqlambda})
for our simple case. However the dependence on $k^{0}_{F}R_{0}$ is somewhat
weaker since the exponent is $2/3$ instead of $1$. 

Our result Eq.(\ref{eqcoll2}) is coherent with what might 
be expected physically. Indeed
this can not be the density right at the center of the trap which is
relevant for the instability. One has rather to consider the average
density over a region of typical size a few $ k_F ^{-1}$. This average
density is lower than the nominal density right at the center. We expect
therefore that the threshold for the instability is raised, compared to
what one could obtain by considering only the density at the center.
This is just what we obtain. Nevertheless the shift of the instability
is rather small since it is of order $ 1/(R_{0}k_{F}^0)^{2/3} $. On the other
hand this is coherent with the fact in the limit $ R_0 \rightarrow \infty$
we have to recover the homogeneous case with no shift at all. Hence
the result has to depend on the ratio between the microscopic length
$ k_F ^{-1}$ and the macroscopic one $ R_0$, wich is small. Therefore
we come to the conclusion that the effect of the trapping potential is
unlikely to be responsible for the lack of collapse found in experiments.
On the other hand it is not completely clear that the very elongated
shape of the traps used in most of 
this kind of experiments would not play some
role. Indeed in this case the above ratio would not be that small.
Nevertheless we have not explored this more complex situation. However in at 
least the ENS experiment \cite{bourdel} the trap is not so far from
isotropic, so strong anisotropy does not hold for explaining the lack
of collapse. 

\subsection{General case}

Finally we generalize our treatment by getting rid of our approximate
evaluation of the static response function
of the system. Indeed we used above the simple RPA to obtain it.
However our above derivation makes it clear that, because of our
semiclassical approximation, all what we need is the response function
of the homogeneous system. Hence we can use the general framework of Fermi
liquid theory to discuss it. Let us recall that this framework is an
exact one, and that it applies to our case of interest, that is a
strongly interacting neutral Fermi system \cite{pn}, 
just as it does for normal and superfluid liquid
$ ^{3}$He. Naturally there is a counterpart to the fact that this
framework is exact, which is that it does not provide the explicit
values of the constants it introduces. However this is not so important
here since our approximate treatment provides us already with
order of magnitude for these constants. On the other hand it is quite
important to know that the only things which are not exact in our
treatment are the values of these constants.

We found above from the RPA that the
response function $ \Pi$ for the interacting system could be obtained
from the response function $ \Pi^0$ of the non interacting system by
$\Pi^{-1}=(\Pi^0)^{-1}-g$. Now in Fermi liquid theory \cite{pn}, for
zero wavevector, the exact value of $ \Pi$ is given by:
\begin{eqnarray}
- \Pi^{-1} = \frac{1}{N_0} + \frac{F_0^s}{N_0}
\end{eqnarray} 
where $N_0= m^* k_F / \pi ^2  $, is the density of states,
with $m^*$ the effective mass, and $F_0^s$ the $ \ell = 0$ symmetric
Landau parameter. We see that we have merely to replace the bare
mass by the effective mass, and replace $g$ by $F_0^s/N_0$ to obtain
the exact result (note that we have a factor 2 difference with 
our above explicit expression for the density of states because we
consider the two spin populations). In particular the collapse 
instability corresponds to the well-known condition $F_0^s=-1$.
Hence we only need to know the exact dependence of $N_0$ and $F_0^s$
as a function of particle density in order to recast our above
calculation under an exact form. More precisely we need only to
know this dependence in the vicinity of the collapse density. Finally
in order to write the generalization of Eq.(\ref{eqdiffairy}) we need
not only our response function for zero wavevector, but also the
lowest correction due to the fact that we work at nonzero wavevector $q$.
For dimensional reasons this correction will be in $(q/k_F)^2$ but the
coefficient will not be given by the free gas result, as we have done
above. More generally this coefficient is beyond the reach of standard
Landau's Fermi liquid theory. This leads us to write finally the small $q$ 
expansion:
\begin{eqnarray}
- \Pi^{-1} = \frac{1}{N_0}(1+b \frac{q^2}{k_F^2}) + \frac{F_0^s}{N_0}
\end{eqnarray}
where we had $b=1/12$ in the free particle case. Hence the framework
we used in our calculation is exact, only the constants which come in
have to be modified. We do not rewrite
here the generalization of our above calculations
since it is straighforward to do it and the results are not expected
to lead to qualitative changes with respect to our above ones.

\section{Conclusion}
In this work, we have considered the zero frequency unstable
density fluctuations for ultracold fermions in a harmonic trap.
More precisely we have studied how quantum effects modify the
corresponding mode. We have used a semiclassical treatment
justified by the fact that the trap is large compared to the
microscopic quantum scale. In a first step we have
treated the interactions within the RPA approximation. We
have then made a fully general analysis within Fermi liquid theory.
Our results show that there is a zero frequency mode, with a size 
large compared to the inverse Fermi wave vector at the center of the trap 
 but small compared to the cloud size. The threshold for instability is 
modified by the trap. However, even though we find a modification
in the threshold instability,  this does not seem to explain the absence of 
instability, as it is observed experimentally. It is interesting to note
that, by contrast, quantum effects have a strong influence on the
corresponding instability for Bose systems. This is basically because
in this case the correspondent of 
our dimensionless parameter $R_{0}k_{F}^0$ becomes of order unity.
It would be of interest
to devise a model allowing to go continuously from the Fermi case
to the Bose case, and see the importance of quantum effects grow
along the way.

* Laboratoire associ\'e au Centre National
de la Recherche Scientifique et aux Universit\'es Paris 6 et Paris 7.


\begin{references}
\bibitem{moden0}
Indeed, one can show that for an hydrodynamic mode of frequency
$\omega$, one always has \mbox{$\omega^2>0$}, except for the "neutral" mode. 
In order to prove this, we consider Eq.(1) of \cite{fuchs} for the
chemical potential fluctuation $w$ :
\mbox{$\nabla\cdot(n_0\,\nabla\,w)+m\omega^2\frac{\partial\,n_0}{\partial\mu}\,w=0$}
where $n_0(\bf r)$ is the equilibrium density, $m$ is the atomic mass and 
$\frac{\partial\,n_0}{\partial\mu}$ is the neutral mode.
We multiply the above equation by $w$ and integrate over space. 
Integration by parts gives a surface term equals to zero, because
$n_0$ is zero on the cloud surface. We then get for the square of the mode frequency,
for any confining potential, in any dimension:
$\omega^2=(1/m)\left[\int d{\bf r}\,n_0\,(\nabla\,w)^2\right]/
\left[\int d{\bf r}\,\frac{\partial n_0}{\partial \mu}\,{w}^2\right]$.
This shows  that $\omega^2 >0$, except for the neutral mode for which $w=1$.
\bibitem{instabmod}More precisely the instability appears when the frequency
has a (infinitely small) positive imaginary part. 
This could in principle happen
with a non zero real part of the frequency. However the standard situation
for the appearance of an instability, as it is found for a Bose condensate
with negative scattering length,
is the case where the square of the
frequency $ \omega ^{2}$ is real and goes from a positive to a negative
value by going through zero, in which case the frequency itself goes from
real to purely imaginary. Hence the instability appears when the frequency
is zero. We assume that we are also in this situation.
\end{references}
\end{document}